\documentclass[12pt]{iopart}

\expandafter\let\csname equation*\endcsname\undefined
\expandafter\let\csname endequation*\endcsname\undefined
\usepackage{amsmath}
\usepackage{amssymb}

\usepackage{dsfont}
\usepackage[dvipdfmx]{graphicx}
\usepackage{siunitx}
\usepackage{floatrow}
\usepackage{caption}
\usepackage{stackrel}
\usepackage[T1]{fontenc}  
\usepackage{lmodern}    

\begin{document}

\title[Measuring Magnetic 1/f Noise in Superconducting Microstructures]{Measuring Magnetic 1/f Noise in Superconducting Microstructures and the Fluctuation-Dissipation Theorem}

\author{
	M~Herbst, 
	A~Fleischmann, 
	D~Hengstler, 
	D~Mazibrada, 
	L~Münch, 
	A~Reifenberger, 
	C~Ständer, and
	C~Enss
}

\address{Kirchhoff Institute for Physics, Heidelberg University, Heidelberg, Germany}
\ead{matthew.herbst@kip.uni-heidelberg.de}

\begin{abstract}

The performance of superconducting devices like qubits, SQUIDs, and particle detectors is often limited by finite coherence times and $1/f$ noise. Various types of slow fluctuators in the Josephson junctions and the passive parts of these superconducting circuits can be the cause, and devices usually suffer from a combination of different noise sources, which are hard to disentangle and therefore hard to eliminate. One contribution is magnetic $1/f$ noise caused by fluctuating magnetic moments of magnetic impurities or dangling bonds in superconducting inductances, surface oxides, insulating oxide layers, and adsorbates. In an effort to further analyze such sources of noise, we have developed an experimental set-up to measure both the complex impedance of superconducting microstructures, and the overall noise picked up by these structures. This allows for important sanity checks by connecting both quantities via the fluctuation-dissipation theorem. Since these two measurements are sensitive to different types of noise, we are able to identify and quantify individual noise sources. The superconducting inductances under investigation form a Wheatstone-like bridge, read out by two independent cross-correlated dc-SQUID read-out chains. The resulting noise resolution lies beneath the quantum limit of the front-end SQUIDs and lets us measure noise caused by just a few ppm of impurities in close-by materials. We present measurements of the insulating $\textrm{SiO}_2$ layers of our devices, and magnetically doped noble metal layers in the vicinity of the pickup coils at $T = 40\,\textrm{mK} - 800\,\textrm{mK}$ and $f = 1\,\textrm{Hz}\ - 100\,\textrm{kHz}$.

\end{abstract}


\maketitle

\section{Introduction}
\label{chap:Introduction}

Excess $1/f$ noise is a ubiquitous feature found in superconducting devices, often limiting their performance or application range. With recent progress in quantum computing \cite{Ladd2010, Arute2019, Wu2021}, $1/f$ noise in superconducting quantum bits (qubits) \cite{Clarke2008} has been a particular focus of research. In charge qubits, 
for instance, it was found that such noise is caused by charged two-level fluctuators inducing charge fluctuations in the qubit \cite{Nakamura2002, Ithier2005}. In devices with Josephson junctions, $1/f$ noise can be caused by fluctuations of the Josephson energy and the critical current originating from two-level fluctuators in the tunneling barrier \cite{Ithier2005, Wakai1986, VanHarlingen2004}.
Poorly understood, however, is excess low frequency magnetic flux noise, which can limit the coherence time of flux-, phase-, and transmon-type qubits and thus inhibit their scaling to more complex systems \cite{Yoshihara2006, Bialczak2007, Paladino2014}.
It is also present in superconducting quantum interference devices (SQUIDs) where it dominates at low frequencies \cite{Kempf2016}, and in detectors such as metallic magnetic calorimeters (MMCs) \cite{Fleischmann2005}.

Magnetic $1/f$ noise is characterized by a power spectral density of $S_\varPhi (f) = A^2\,(f /\textrm{Hz})^{-\alpha}$, with a largely device independent slope $\alpha \lesssim 1$ and an amplitude $A$ in the order of a few $\text\textmu\Phi_0/\sqrt{\textrm{Hz}}$, where $\Phi_0 = \frac{h}{2\,e}$ is the magnetic flux quantum.
Its microscopic origin is not fully understood. In MMCs, $1/f$ noise scales with the amount of magnetic moments in the sensor material, suggesting that it originates from the paramagnetic, erbium-doped gold or silver sensor \cite{Fleischmann2009}. For SQUIDs, flux, and phase qubits, different models consider magnetic flux noise from the stochastic hopping of electrons \cite{Koch2007}, paramagnetic dangling bonds \cite{deSousa2007}, strongly interacting surface spins \cite{Faoro2008, Faoro2012}, metal-induced gap states \cite{Choi2009, Szczesniak2021}, or adsorbed $\textrm O_2$ molecules \cite{Wang2015, Kumar2016}.
Both the experimental verification of theories and the reduction of noise in specific set-ups can be difficult, since devices usually suffer from a combination of different noise sources, which are hard to disentangle and therefore hard to eliminate.

We have taken this problem as motivation and have developed a device to measure magnetic flux noise of superconducting circuits, with the focus lying on isolating individual noise sources.
The general idea is to combine two different methods of measuring noise in the same device: 
First, a direct measurement of noise via a pair of two-stage dc-SQUID read-out chains, which are cross-correlated in order to reduce read-out noise. This will give us a measure of the sum of all noise originating from the superconducting circuit. 
Second, a measurement of the complex susceptibility of inductances in close proximity to a sample material. By applying the fluctuation-dissipation theorem, we then calculate the magnetic flux noise caused by the sample.
A comparison of results from the two measurement modes tells us which noise components originate from the sample material and which from other parts of the system.

In the following, we present our experimental set-up, including a detailed explanation of the two measurement modes. There follows measurements of a highly paramagnetic sample material with a known amount of localized magnetic moments, as well as sputtered $\textrm{SiO}_2$, an insulator commonly used in superconducting microstructures. 


\section{Experimental Methods}
\label{chap:Experimental Methods}

Figure~\ref{fig:device_shematic} shows a sketch of the set-up. The central part is a Wheatstone-like bridge consisting of four almost identical, superconducting, microstructured, meander-shaped coils placed on a $3\,\textrm{mm} \times 3\,\textrm{mm}$ silicon chip. Each meander consists of 50 lines with a width of $5\,\text\textmu\textrm{m}$ and a pitch of $10\,\text\textmu\textrm{m}$, resulting in an inductance of $L_0 = 7 \,\textrm{nH}$. Two opposite meander coils are coated with the sample material (orange), changing their inductance to $L = L_0(1 + \chi F)$, where $\chi$ is the magnetic susceptibility of the sample material, and $F$ is the filling factor. The sample is galvanically decoupled from the coils by a thin layer of $\textrm{SiO}_2$ covering the entire chip. Further components connected to the main chip allow us to read out the generated noise in two different measurement modes.

\noindent\begin{minipage}[tbp]{.5\textwidth}
	\centering
	\includegraphics[width=\linewidth, keepaspectratio]{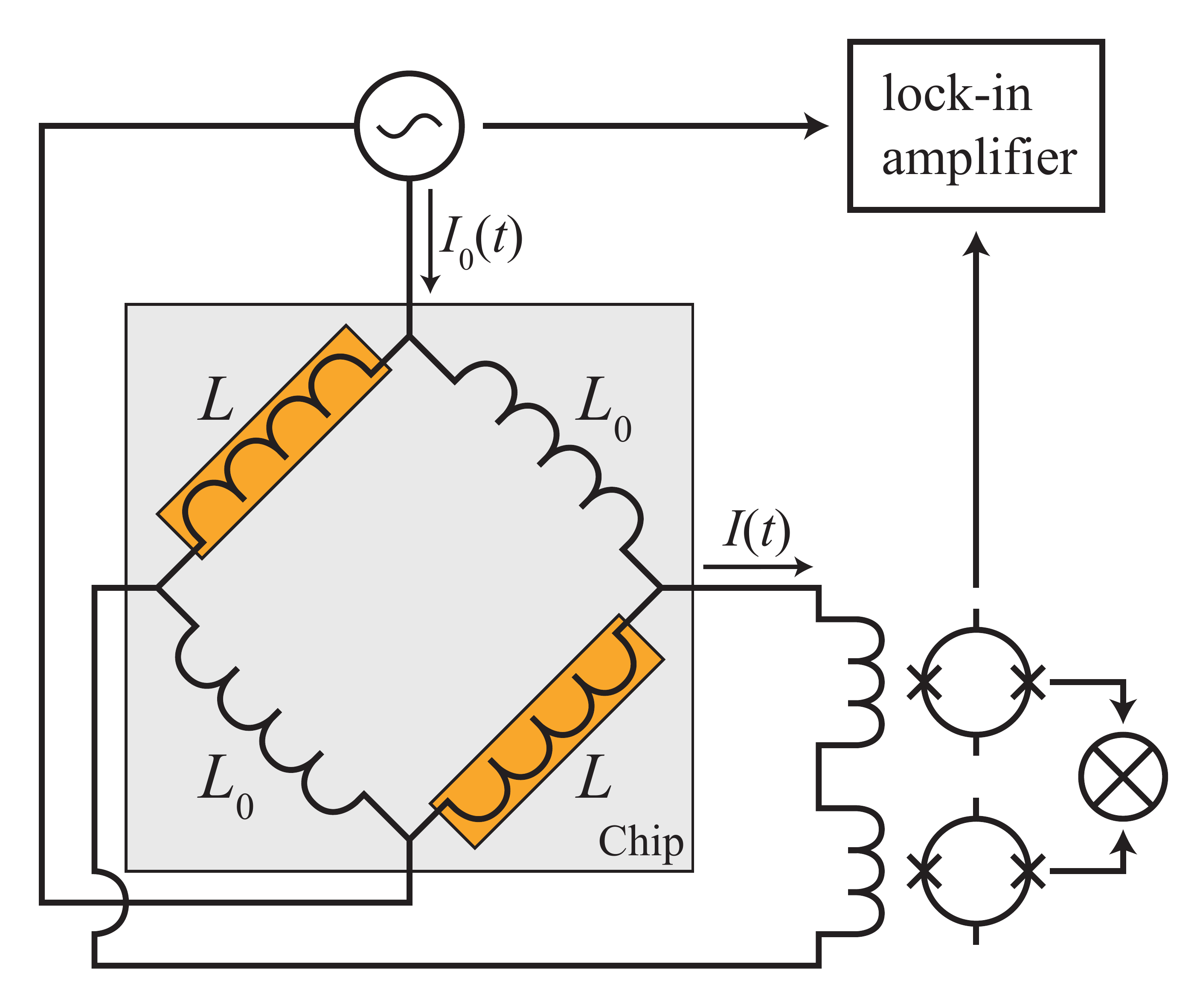}
\end{minipage}\hfill
\begin{minipage}[b]{.45\textwidth}
	\captionof{figure}{Circuit diagram of the set-up with sample material (orange) placed on opposite inductors. Noise can be read out directly using a pair of cross-correlated two-stage dc-SQUID read-out chains. Alternatively, we can measure the susceptibility of the sample material by applying an AC signal $I_0(t)$ and determining the amplitude and phase shift of the current $I(t)$ using a lock-in amplifier.}
	\label{fig:device_shematic}
	\vspace{-3.5cm}
\end{minipage}


\subsection{Cross-correlated noise read-out}
\label{chap:cc}
In cross-correlation mode, we read out the noise $a(t)$ of the superconducting microstructure via a pair of two-stage dc-SQUID read-out chains, where the input coils of the two chains are connected in series (see figure~\ref{fig:device_shematic}). From the read-out chains originate additional undesired noise contributions $n_1(t)$ and $n_2(t)$. In a realistic experimental setting, there is a finite amount of cross-talk between the read-out chains. We model this cross-talk from the respective other channel as a contribution with relative fraction~$\delta$.
As a result, the actually measured time traces $s_1(t)$ and $s_2(t)$ are

\begin{equation}
	\begin{aligned}
		s_1(t) &= (1-\delta) \cdot(a(t) + n_1(t)) + \delta \cdot (a(t) + n_2(t)) \quad \textrm{and} \\
		s_2(t) &= (1-\delta) \cdot(a(t) + n_2(t)) + \delta \cdot(a(t) + n_1(t)) \quad.
	\end{aligned}
\end{equation}
Solving the system of equations results in the time traces
\begin{equation}
	\label{eq:uv}
	\begin{aligned}
		u(t) = {a}(t) + {n}_1(t) &= \frac{{s}_1(t) - \delta({s}_1(t) + {s}_2(t))}{1 - 2 \delta} \quad \textrm{and} \\
		v(t) = {a}(t) + {n}_2(t) &= \frac{{s}_2(t) - \delta({s}_1(t) + {s}_2(t))}{1 - 2 \delta} \quad,
	\end{aligned}
\end{equation}
which have reduced cross-talk.
As a next step, we cross-correlate these two time traces \cite{Rubiola2010}. The
cross-correlated power spectral density $S_{uv}(f)$ is defined as the real part $\textrm{Re}\left\{\,\right\}$ of the Fourier transform $\mathcal F$ of the cross-correlation function
\begin{equation}
	\begin{aligned}
		R_{uv}(\tau) 
		& = \mathds E [u(t)\,{v(t-\tau)}] \\
		& = \lim_{T \to \infty} \frac{1}{T} \int_{-T/2}^{T/2} u(t)\,{v(t-\tau)} \textrm dt \\
		& = \lim_{T \to \infty} \frac{1}{T} \int_{-\infty}^{\infty}  u_T(t)\,{v_T(t-\tau)} \textrm dt \quad .
	\end{aligned}
\end{equation}
Here, $\mathds E[\,\,\,]$ is the expectation value and the subscript $T$ represents a truncation to the measurement time $T$. 
Using the convolution theorem, we find
\begin{equation}
		S_{uv}(f) =
		\textrm{Re} \left\{\mathcal F[R_{uv}(\tau)] \right\} = 
		\lim_{T \to \infty} \frac{1}{T} \,
		\textrm{Re} \left\{
			 \mathcal F^*[u_T(t)](f) \, \mathcal F[v_T(t)](f) 
		\right\}\quad ,
\end{equation}
where the asterisk denotes the complex conjugate.
In order to simplify notation, we introduce the Fourier transform $X = \mathcal F[x_T(t)]$ of a time trace $x(t)$ truncated to $T$. In an experimental setting where we average over $N$ measurements, each with a finite measurement time $T$, the averaged cross-correlated power spectral density then is
\begin{equation}
	\begin{aligned}
		\langle S_{uv}\rangle_N
		& = \frac{1}{T}\, \textrm{Re}\left\{\langle  U^* V               \rangle_N \right\} \\
		& = \frac{1}{T}\, \textrm{Re}\left\{\langle (A^* + N_1^*) (A+N_2)\rangle_N \right\} \\
		& = \frac{1}{T}\, \textrm{Re}\left\{\langle  A^*A                \rangle_N
							+ \langle  A^*N_2    \rangle_N
							+ \langle N_1^*A   \rangle_N
							+ \langle N_1^*N_2 \rangle_N
							\right\} \\
		& = \frac{1}{T}\, \textrm{Re}\left\{\langle A^*A \rangle_N 
							+ \mathcal O \left(1/\sqrt N \right)
							\right\} \quad .
	\end{aligned}
\end{equation}
For sufficiently large $N$, the statistically independent terms vanish and we are left with the cross-talk corrected power spectral density $S_a = \frac{1}{T}\, \textrm{Re}\left\{ \langle A^*A \rangle_N \right\}$ of only the noise in our superconducting microstructure. Note that while read-out noise $n_1$ and $n_2$ is removed, $S_a$ might still have contributions other than the magnetic flux noise caused by the sample material, such as Johnson noise or magnetic flux noise originating from other sources within the input circuit.

\begin{figure}[tbp]
	\begin{center}
		\includegraphics[width=1\linewidth, keepaspectratio]{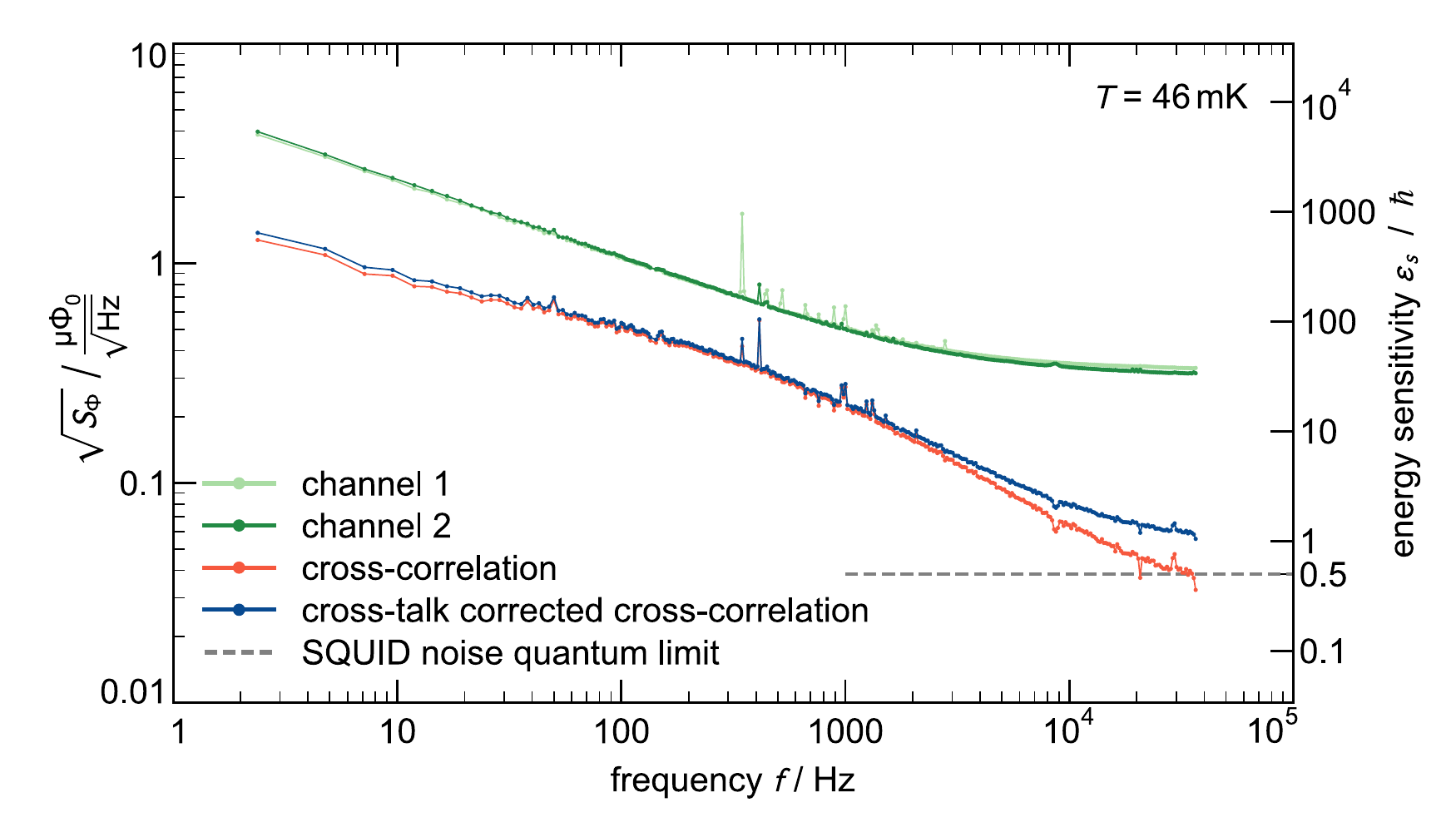}
		\caption{Noise spectra of an $\textrm{SiO}_2$ sample taken in cross-correlation mode. Data from a single channel appears in green, cross-correlated data in red, and data after cross-talk correction in blue. Via the cross-correlation, SQUID read-out noise is greatly reduced. Since cross-talk is negative, correcting for it increases the noise level.}
		\label{fig:CC mode}
	\end{center}
\end{figure}


We demonstrate this process on data taken from noise measurements of an $\textrm{SiO}_2$ sample with an unknown amount of magnetic impurities. Data of the two individual channels appear in green in figure~\ref{fig:CC mode}. At high frequencies, the spectra are dominated by the well understood white SQUID noise \cite{Tesche1977} with an amplitude of $\sqrt{S_\textrm{s,w}} = 0.29\,\text\textmu\Phi_0/\sqrt\textrm{Hz}$. 
This corresponds to an energy sensitivity $\epsilon_\textrm s = \frac{S_\varPhi}{2\,L_\textrm s}$ of the SQUID of around $28\,\hbar$, where we estimate the SQUID inductance to be $L_\textrm s = 60\,\textrm{pH}$. 
At low frequencies, we observe $1/f$ flux noise 
$S_{\varPhi_\textrm{s}} (f) = S_{\varPhi_\textrm{s}}(1\,\textrm{Hz})\,(f /\textrm{Hz})^{-\alpha}$ with 
$\sqrt{S_{\varPhi_\textrm{s}}(1\,\textrm{Hz})} = 5.5\,\text\textmu \Phi_0/\sqrt{\textrm{Hz}}$ and $\alpha = 0.75$. Both components are significantly reduced after cross-correlation (red), and the underlying noise spectrum of the input circuit and Wheatstone bridge becomes visible. 
Based on the circuit design, we find that the cross-talk must be negative and at most $2\,\%$. We thus choose a value of $\delta = (-1\pm1)\,\%$ (blue).
Since $\delta < 0$, the noise level increases, but the data nevertheless approach the quantum noise limit $\epsilon_\textrm s \gtrsim \hbar/2$ of a single SQUID at $T = 0\,\textrm K$ \cite{Danilov1983}. 
The uncertainty in the cross-talk can result in a systematic error at low noise levels, but the amount of cross-talk between channels can be further reduced by increasing $L_0$. To understand this, consider a magnetic flux change $\partial \varPhi_\textrm m$ in one of the meander coils due to magnetic flux noise of the sample. It is proportional to $L_0$, assuming the increase in $L_0$ is achieved through an increase in the meander footprint and thus also in area covered by the sample. Using Kirchhoff's rules and approximating $L \approx L_0$, 
we find that for our circuit, $\partial \varPhi_\textrm m$ creates a flux change
\begin{equation}
	\partial \varPhi_\textrm{s,m} = \frac{k\sqrt{L_\textrm i L_\textrm s}}{2(L_0 + 2(L_\textrm i + L_\textrm w))}
	\,\partial \varPhi_\textrm m
\end{equation}
in one of the front-end SQUIDs \cite{Clarke2004}. Here, $k$ is the dimensionless coupling constant between SQUID input coil $L_\textrm i$ and the inductance $L_\textrm s$ of the front-end SQUID, and $L_\textrm w$ is the inductance of the aluminum bond wires connecting the Wheatstone bridge with the input coils of the SQUID. At the same time, back-action flux $\partial \varPhi_\textrm x$ induced in the input coil by one of the SQUIDs leads to cross-talk
\begin{equation}
	\partial \varPhi_\textrm{s,x} = \frac{k\sqrt{L_\textrm i L_\textrm s}}{L_0 + 2(L_\textrm i + L_\textrm w)}
	\,\partial \varPhi_\textrm x	
\end{equation}
in the other SQUID. Importantly, $\partial\varPhi_\textrm x$ is dominated by read-out noise and thus independent of $L_0$ to first order. As a result, $\partial \varPhi_\textrm{s,x}$ vanishes for large $L_0$, while $\partial \varPhi_\textrm{s,m}$ stays constant and we find
\begin{equation}
	\frac{\partial \varPhi_\textrm{s,m}}{\partial \varPhi_\textrm{s,x}} \propto L_0 \quad .
\end{equation}
Instead, for experiments where a large $L_0$ is impossible and noise from a fixed $L_0 \ll L_\textrm i$ should be measured, it is best to connect the two SQUID input coils in parallel. For that geometry, we find
\begin{equation}
	\begin{alignedat}{4}
		&\partial \varPhi_\textrm{s,m,par} &
		&= \frac{k\sqrt{L_\textrm i L_\textrm s}}{\,2\,(2L_0 \,+\, (L_\textrm i + L_\textrm w))}&
		&\,\partial \varPhi_\textrm m&
		&\gg \partial \varPhi_\textrm{s,m}\quad \textrm{and} \\
		&\partial \varPhi_\textrm{s,x,par} &
		&= \frac{k\sqrt{L_\textrm i L_\textrm s}}{\frac{(L_\textrm i + L_\textrm w)^2}{L_0} + 	2(L_\textrm i + L_\textrm w)} &
		&\,\partial \varPhi_\textrm x &
		&\ll \partial \varPhi_\textrm{s,x} \quad .
	\end{alignedat}
\end{equation}


\subsection{Noise read-out via susceptibility}
Our device allows for a second measurement mode, where we operate it as a Wheatstone-like bridge by applying an AC signal $I_0(f, t) = \hat{I}_0\,\textrm{sin}(2\pi f t)$ over one diagonal and measuring the current $I(f, t) = \hat{I}\,\textrm{sin}(2\pi f t-\theta)$ flowing across the other diagonal via one of the dc-SQUIDs. A lock-in amplifier provides a precise measurement of the relative amplitude $\hat{I} / \hat{I}_0$ and the phase shift $\theta$ of the signal. 
The latter is caused in part by finite signal transmission speed, which we must correct for, and in part by magnetic moments in the sample material not following the magnetic field caused by $I_0$ instantaneously - a fact we express by attributing the sample material a complex magnetic susceptibility $\chi = \chi' + i\chi''$.
Using Kirchhoff's rules, we find the relation
\begin{equation}
	I(f,t) = I_0(f,t) \frac{L-L_0}{2 \, L_\textrm{tot}} = I_0(f,t)\frac{L_0}{2 \, L_\textrm{tot}} \left(\chi'F + i\chi''F\right) \quad,
\end{equation}
where $L_\textrm {tot} = L_\textrm b + 2\,(L_\textrm i + L_\textrm w)$ is the total inductance of the circuit, containing the inductance $L_\textrm b = \frac{1}{2}(L_0+L)$ of the Wheatstone bridge.
It follows that
\begin{equation}
	\chi'F  = \frac{2\,L_\textrm{tot}}{L_0}\frac{\hat{I}}{\hat{I}_0}\,\textrm{cos}(\theta) 
	\quad \textrm{and} \quad 
	\chi''F = \frac{2\,L_\textrm{tot}}{L_0}\frac{\hat{I}}{\hat{I}_0}\,\textrm{sin}(\theta)\quad.
\end{equation}
When measuring paramagnetic samples, $\chi' F$ should vanish in the limit of low frequencies and high temperatures. In practice, however, this is not the case, since the four coils of the Wheatstone bridge have slightly different inductances due to fabrication inaccuracies. In our model we account for this by replacing one of the base inductors $L_0$ with a slightly altered inductor $L_{0,\textrm a}$. During data evaluation, we then determine $L_{0,\textrm a}$ by extrapolating data to $1/T \rightarrow 0$ using Curie's law and then choosing $L_{0,\textrm a}$ so that $\chi' F(f\rightarrow 0,1/T \rightarrow 0) = 0$. Measurements of three different devices resulted in asymmetries $|L_0 - L_{0,\textrm a}|/L_0$ of less than $0.3\,\%$.

Consider now the imaginary part of the susceptibility. Any non-zero value for $\chi''$ leads to a dissipative component $\textrm{Re}(Z_\textrm b)$ of the Wheatstone bridge's impedance
\begin{equation}
	\begin{aligned}
		\textrm{Re}(Z_\textrm b) 
		&= \textrm{Re}\left(i \, 2\pi f \, L_\textrm b\right) 
		= \textrm{Re}\left(i \,  \pi f \, (L_0 + L)  \right) \\
		&= \textrm{Re}\left(i \,  \pi f \, (2L_0 + L_0(\chi' + i\chi'')F)\right)
		= -\pi f L_0 \chi''F \quad,
	\end{aligned}
\end{equation}
which is positive, since $\chi''<0$.
According to the fluctuation-dissipation theorem, we expect a corresponding voltage noise $S_U(f) = 4k_\textrm BT\,\textrm{Re}(Z_\textrm{W})$ across the impedance $Z_\textrm{tot} = i\,2\pi f L_\textrm{tot}$ of the circuit, resulting in flux noise
\begin{equation}
	\sqrt{S_\varPhi(f)} 
	= k\sqrt{L_\textrm i L_\textrm s}\, \frac{\sqrt{S_U(f)}}{|Z_\textrm{tot}|} 
	= \frac{k\sqrt{L_\textrm i L_\textrm s}}{L_\textrm{tot}}\, \sqrt{\frac{k_\textrm BTL_0|\chi''|F}{\pi f }}
\end{equation}
in the SQUID.
Note that in contrast to noise read out in cross-correlation mode, only asymmetrically distributed magnetic flux noise sources are probed by measuring $\chi''$.


\section{Results and Discussion}
\subsection{Measurement mode comparison}
\begin{figure}[tbp]
	\begin{center}
		\includegraphics[width=1\linewidth, keepaspectratio]{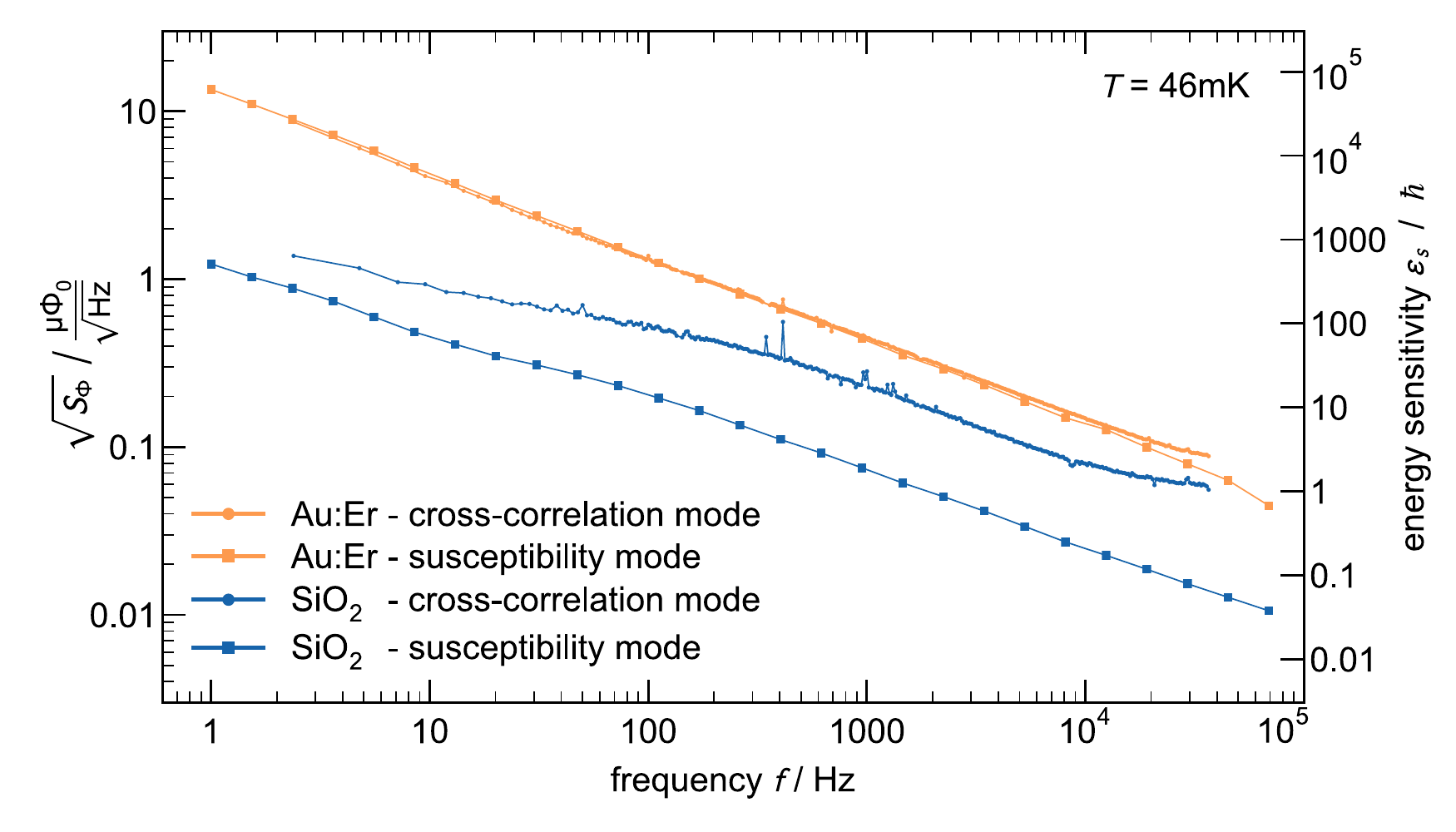}
		\caption{Results of noise measurements on gold doped with $2480\,\textrm{ppm}$ erbium (orange), and on $\textrm{SiO}_2$ with unknown magnetic impurities (blue) at $T = 46\,\textrm{mK}$. Data from susceptibility mode (squares) represents only the magnetic flux noise caused by the respective sample, while data from cross-correlation mode (circles) includes additional components from Johnson noise, white noise, and other 1/$f$ noise.}
		\label{fig:firstresults}
	\end{center}
\end{figure}

As a first demonstration of the functionality of our device, we present measurements from a chip onto which we deposited a $2.58\,\text\textmu\textrm{m}$ thick Au:Er film with an atomic erbium concentration of $x_\textrm{Er} = (2480 \pm 20)\,\textrm{ppm}$.
The localized 4f electrons of the erbium have an effective spin $\tilde S = 1/2$ and an effective $\tilde g = 6.8$, with the gold acting as a passive dilutant to prevent magnetic ordering \cite{Abragam1970}. This results in a system of mostly free magnetic moments, leading to a strongly paramagnetic material with a Curie-Weiss-type temperature dependence of $\chi'$ \cite{Fleischmann2000}.
Data using both measurement modes appear in orange in figure \ref{fig:firstresults}. To first order, the noise is independent of measurement mode, which tells us that almost all of the noise in the superconducting circuit consists of magnetic flux noise caused by the sample. Given the magnetic nature of the erbium, this is not surprising. 
We describe the noise as a $1/f$-type noise $S_{\varPhi,\textrm{Er}} (f) = S_{\varPhi,\textrm{Er}}(1\,\textrm{Hz})\,(f /\textrm{Hz})^{-\alpha}$ with $S_{\varPhi,\textrm{Er}}(1\,\textrm{Hz}) = \left(13.1\,\text\textmu \Phi_0/\sqrt{\textrm{Hz}}\right)^2$ and $\alpha = 1.00\pm0.03$. 
Using finite element simulations of the magnetic field distribution in the material, we are able to assign a value of $S_\textrm{Er}(1\,\textrm{Hz}) = (0.115\pm0.006)\,\mu_\textrm B^2/\textrm{Hz}$ per erbium atom \cite{Fleischmann2005}.
Our result is compatible with noise that was previously observed in MMCs containing erbium alloys, where a value of $\sim 0.1\,\mu_\textrm B^2/\textrm{Hz}$ per erbium atom was estimated with larger uncertainties \cite{Fleischmann2009}.

On a second device we deposited a $440\,\textrm{nm}$ thick layer of $\textrm{SiO}_2$ over the entire device, with two square holes in the $\textrm{SiO}_2$ over opposite meanders giving us the required asymmetry. 
The magnetic flux noise measured using susceptibility mode appears as blue squares in figure~\ref{fig:firstresults}. It has a slightly frequency dependent slope of about $f^{-0.9}$ and with 
$S_{\varPhi,\textrm{SiO}_2}(1\,\textrm{Hz}) = \left(1.65\,\text\textmu \Phi_0/\sqrt{\textrm{Hz}}\right)^2$, 
it lies significantly below the total noise of the circuit, as measured in cross-correlation mode (blue dots). Other noise components are dominant in this set-up, which we analyze in section~\ref{chap:temperature dependence}. 
Comparing the magnetic flux noise of both samples, we estimate that roughly $60\,\textrm{ppm}$ of impurities are magnetically active at $1\,\textrm{Hz}$, assuming that the impurities in $\textrm{SiO}_2$ have a similar magnetic moment to erbium. More likely, however, is that we are mostly measuring dangling bonds with a spin of $1/2$ and a $g$-factor of $2$ with an appropriately scaled concentration of around $200\,\textrm{ppm}$. From calibration measurements, we determine that this value lies around a factor of three above the resolution limit of susceptibility mode.

\subsection{Temperature dependence}
\label{chap:temperature dependence}
\begin{figure}[tbp]
	\begin{center}
		\includegraphics[width=0.96\linewidth, keepaspectratio]{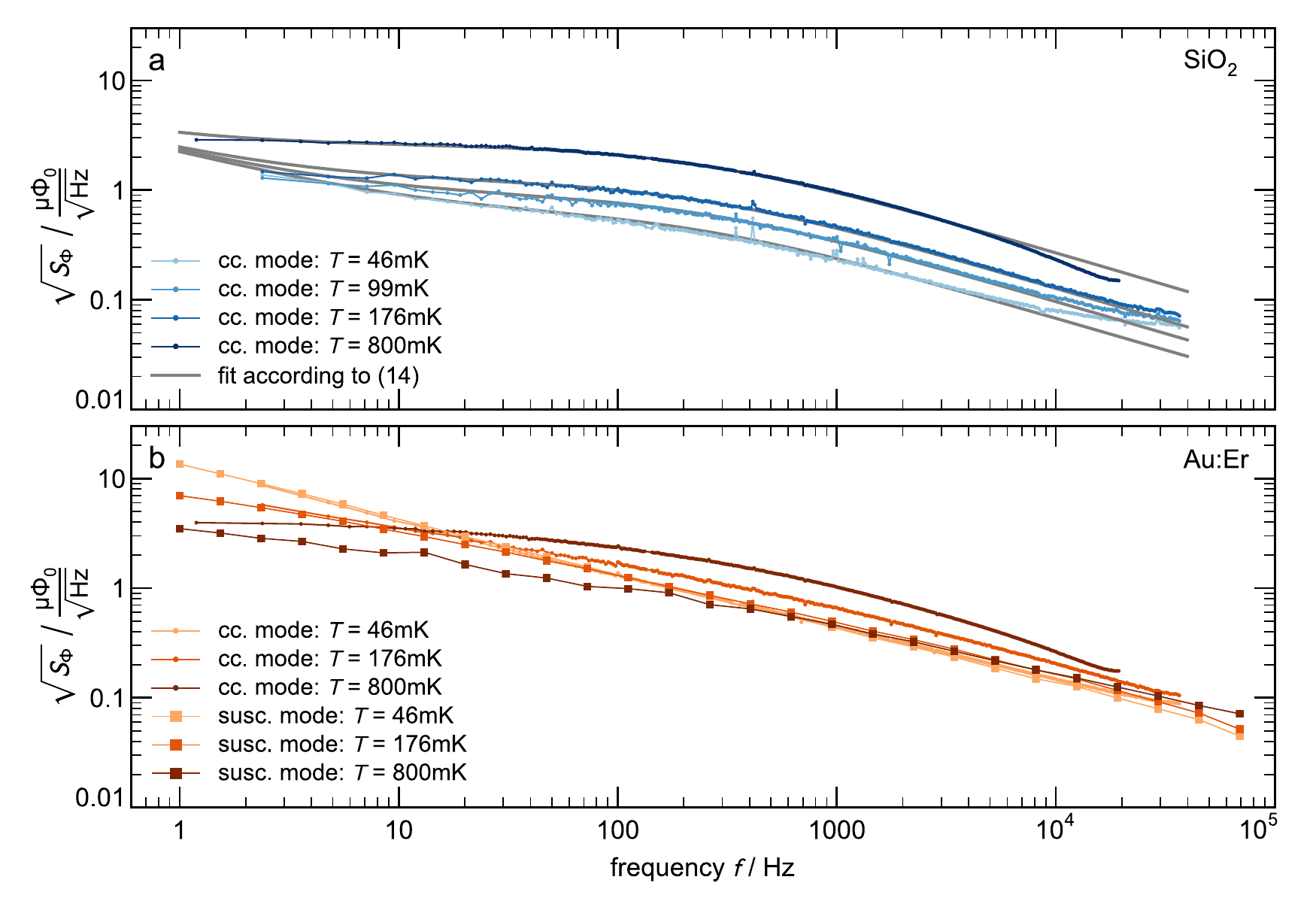}
		\caption{\textbf{a)} Data from noise measurements of $\textrm{SiO}_2$ in cross-correlation mode at different temperatures. A fit given by (\ref{eq:Johnsonfit}), which is based on a combination of magnetic Josephson noise and a $1/f$ component, matches data well. \textbf{b)} Temperature dependent noise of Au:Er for both measurement modes. Johnson noise also affects cross-correlated data at highest temperatures, while the magnetic flux noise is largely temperature independent.}
		\label{fig:T dependence}
	\end{center}
\end{figure}

We gain further insight into the origin of different noise sources by repeating the above measurement at different temperatures. Cross-correlated data of $\textrm{SiO}_2$ appears in figure~\ref{fig:T dependence}a for temperatures between $46\,\textrm{mK}$ and $800\,\textrm{mK}$. 

We observe a general temperature dependence of the form $\sqrt{S}\propto\sqrt{T}$ over large parts of the spectrum, which we describe with the fit
\begin{equation}
	\sqrt{S_{\varPhi,\textrm{fit}}(f, T)} = \sqrt{S_\textrm{J}\,\frac{T}{1\,\textrm{K}} \left(1 + \frac{2\,f}{\pi f_\textrm c}\right)^{-\alpha} + S_{1/f} / f}\quad ,
	\label{eq:Johnsonfit}
\end{equation}
which appears in gray in figure~\ref{fig:T dependence} and matches data well for all but the highest frequencies. The fit function consists of two parts: First, a component with amplitude $S_\textrm{J} = (2.9\,\text\textmu\Phi_0/\sqrt{\textrm{Hz}})^2$, which we attribute to thermal motion of electrons (Johnson noise) in the copper experimental platform coupling magnetically into the loop formed by the bond wires between our device and the front-end SQUIDs. Due to the skin effect, there is a low-pass behavior with $f_\textrm c = 143\,\textrm{Hz}$ and $\alpha = 1.19$. Such magnetic Johnson noise has been previously observed in similar set-ups \cite{Beyer2007}. The second part is a $1/f$ component with 
$S_{1/f} = \left(2.16\,\text\textmu\Phi_0/\sqrt{\textrm{Hz}}\right)^2$, which is caused at least in part by magnetic flux noise from the $\textrm{SiO}_2$, as measurements in figure~\ref{fig:firstresults} show. Note that the susceptibility mode measurement in figure~\ref{fig:firstresults} is based on holes in the $\textrm{SiO}_2$ layer, so only noise caused by these asymmetric sections is measured. In cross-correlation mode, however, magnetic flux noise from $\textrm{SiO}_2$ on other parts of the chip can also contribute, such as the lines leading to the bond pads, or even from $\textrm{SiO}_2$ on the front-end SQUID chips. 
As a result, we are not surprised that the $1/f$ component from cross-correlation mode lies above the noise measured using susceptibility mode.

Data of the Au:Er sample appear in figure~\ref{fig:T dependence}b for both measurement modes. The magnetic flux noise (squares) is largely temperature independent, which matches previous observations \cite{Fleischmann2009}. At highest temperatures, however, we observe reduced noise at low frequencies, since the variable inductance of the aluminum bond wires near $T_\textrm c$ influences the measurement. Data from cross-correlation mode show an additional component at high temperatures and medium frequencies, which is well described by equation (\ref{eq:Johnsonfit}). This points toward that noise component being sample independent and originating from the experimental set-up, reaffirming the assumption that the source is magnetic Johnson noise.


\subsection{Susceptibility measurements}

\begin{figure}[tbp]
	\begin{center}
		\includegraphics[width=1\linewidth, keepaspectratio]{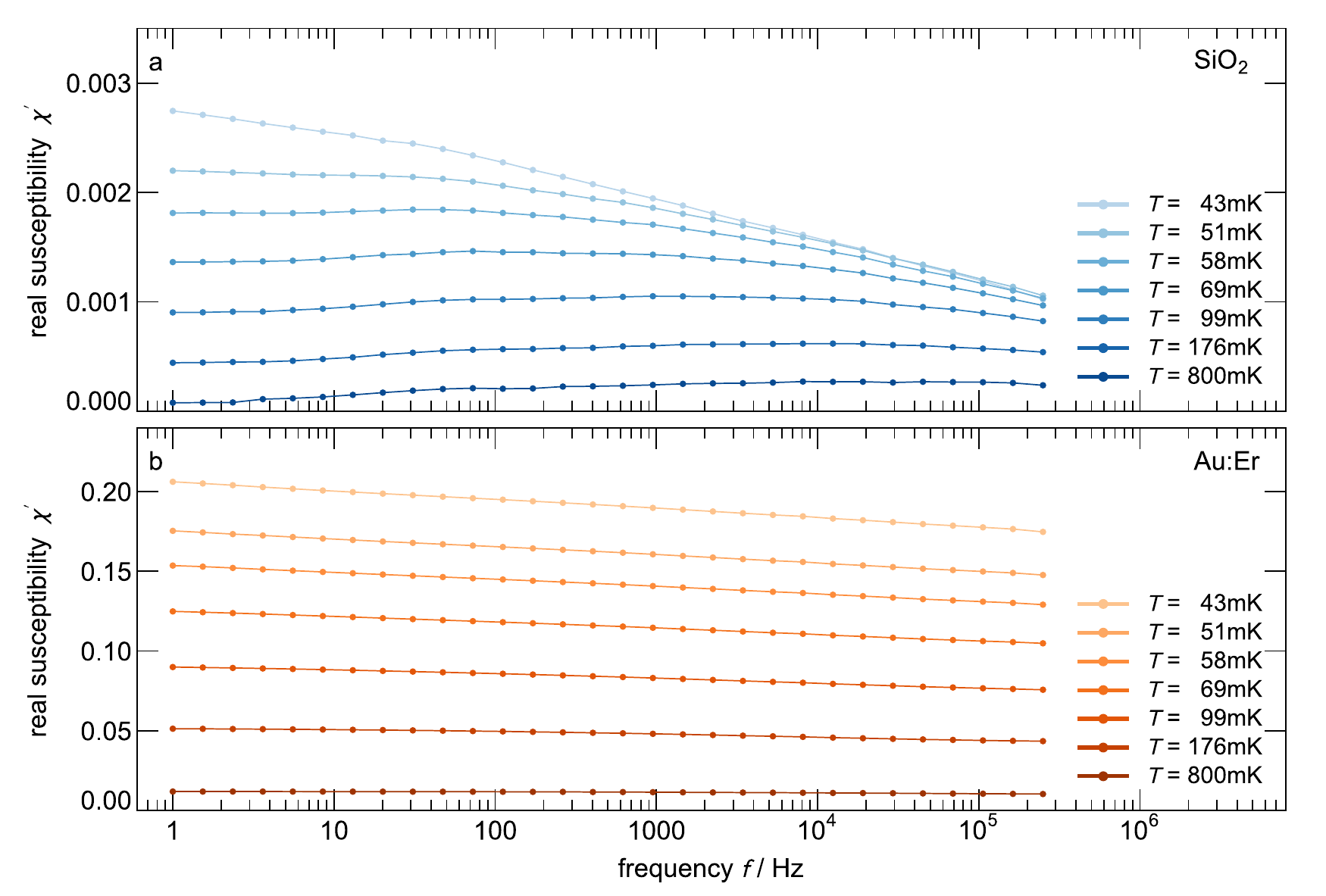}
		\caption{Data of the real part of the AC susceptibility appears as a function of frequency and temperature for $\textrm{SiO}_2$ (\textbf{a}) and Au:Er (\textbf{b}). For individual frequencies, a Curie-Weiss law describes data well, while for individual temperatures a frequency dependence is visible.}
		\label{fig:realpart}
	\end{center}
\end{figure}

Besides extracting the magnetic flux noise from $\chi''$ via the fluctuation-dissipation theorem, we can also gain insight on the dynamics of magnetic moments in the sample material from the real part $\chi'$ of the complex AC susceptibility. Figure~\ref{fig:realpart} shows the frequency dependence of the corresponding data for $\textrm{SiO}_2$ (a) and Au:Er (b). When considering data for either sample at a fixed frequency, $\chi'$ increases towards lower temperatures and saturates in a Curie-Weiss like fashion, as is expected for paramagnetic samples. For fixed temperatures, we observe a general trend of decreasing susceptibility with increasing frequency. To understand this, we model each magnetic moment $\mu_i$ as a Debye relaxator with a relaxation time $\tau_i$, so that at an angular frequency~$\omega$, it contributes with a fraction $\left(1+(\omega\tau_i)^2\right)^{-1}$ of its magnetic moment to $\chi'$.
Averaging over many relaxators with different relaxation times thus yields different shapes for the averaged $\chi'(f)$ curve, depending on the distribution of relaxation times. In our case, for instance, we note that as the frequency increases, $\chi'(f)$ drops quickly for the glassy $\textrm{SiO}_2$ in which slow phononic relaxation with a long $\tau_i$ dominates. In Au:Er, with fast electron-spin relaxation times, magnetic moments can have shorter relaxation times and the frequency dependence of the susceptibility is smaller. In fact, Au:Er data for $T\leq69\,\textrm{mK}$ are well described by $\chi'\propto-\textrm{log}(f)$, telling us that relaxation times must be spaced logarithmically. Using
\begin{equation}
	\label{eq:Lun81}
	\chi'' = \frac{\pi}{2}\frac{\partial \chi'}{\partial \textrm{ln}f}
\end{equation}
derived for spin-glass systems with their broad distribution of relaxation times \cite{Lundgren1981}, we get a frequency independent loss $\chi''$ and via the fluctuation-dissipation theorem once again the $1/f$ noise presented above. Calculating noise in this way is consistent with the direct calculation via the phase shift of the AC signal, but has larger uncertainties. It does show, however, that (\ref{eq:Lun81}) is applicable for our Au:Er sample, which is remarkable, since we expect the spin-glass transition temperature of our sample to lie in the low mK range \cite{Fleischmann2005}. A broad distribution of relaxation times in Au:Er could originate from the complex magnetic nature of the material, which includes RKKY and dipole-dipole interactions between erbium ions, as well as crystal field effects from the gold lattice \cite{Fleischmann2000}. Simulations of magnetically similar Ag:Er alloys have shown that these interactions can lead to a wide range of energy level spacing, with individual pairs or clusters of erbium atoms having energy gaps of well above $100\,\textrm{mK} \cdot k_\textrm B$ \cite{Herbst2022}.


\section{Conclusion}

We have presented a novel device which allows us to analyze noise sources in superconducting microstructures. By combining cross-correlated read-out and AC susceptibility measurements, we are able to distinguish and quantify SQUID noise, magnetic Johnson noise, and magnetic $1/f$ noise. Our device is thus a tool to map out potential noise sources in an experimental set-up before performing the actual experiment. This may be applied to a wide range of superconducting microstructures.
Thinkable, for instance, is replacing the meander-shaped coils with SQUIDs or qubits, in order to measure the magnetic flux of an entire device. 

Besides disentangling noise sources in an experimental set-up, we are also able to analyze the magnetic nature of sample materials. In measurements on Au:Er, we find a nearly temperature independent $1/f^\alpha$  magnetic flux noise component  with an amplitude of $S_\textrm{Er}(1\,\textrm{Hz}) = 0.115\,\mu_\textrm B^2/\textrm{Hz}$ per erbium ion and $\alpha = 1.00$. We are able to measure noise beneath the quantum limit, enabling us to quantify the $1/f$ noise of weakly magnetic $\textrm{SiO}_2$. By combining these results with measurements of the temperature and frequency dependence of the real part of our samples' susceptibility, 
we obtain a detailed picture of the dynamics of the magnetic moments.

\section{Data Availability}
The datasets generated during and/or analyzed during the current study are available in the Zenodo repository doi.org/10.5281/zenodo.7993965. \\

\section{Conflict of Interest}
The authors declare that they have no conflict of interest.

\section{Acknowledgements}
The research leading to these results has received funding from the European Union’s Horizon 2020 Research and Innovation Programme, under Grant Agreement no 824109 (European Microkelvin Platform).
This work has been performed with support of the research project SuperLSI (reference number: 13N16255) which is funded by the German ministry of education and research (BMBF) within the research program `Enabling Technologies für die Quantentechnologien'.

\section{References}

\bibliographystyle{unsrt_edited}
\bibliography{library.bib}

\end{document}